\documentclass{article}

\usepackage{PRIMEarxiv}

\usepackage[utf8]{inputenc} 
\usepackage[T1]{fontenc}    
\usepackage{hyperref}       
\usepackage{url}            
\usepackage{booktabs}       
\usepackage{amsfonts}       
\usepackage{nicefrac}       
\usepackage{microtype}      
\usepackage{lipsum}
\usepackage{fancyhdr}       
\usepackage{graphicx}       
\usepackage{amsmath}
\usepackage{algorithm}
\usepackage{algorithmic}
\usepackage{multirow} 
\graphicspath{{media/}}     
\usepackage{hyperref}
\usepackage{float}
\DeclareUnicodeCharacter{0650}{}
\title{DKD-KAN: A Lightweight knowledge-distilled KAN intrusion detection framework, based on MLP and KAN}

 \author{
   Mohammad Alikhani \\
   Faculty of Electrical Engineering \\
   K.N. Toosi University of Technology \\
   Tehran, Iran\\
   \texttt{m.alikhani2@email.kntu.ac.ir} \\
 }

\begin{document}
\maketitle

\begin{abstract}
Cyber-security systems often operate in resource-contained environments, such as edge environments and real-time monitoring systems, where model size and inference time are crucial. A light-weight intrusion detection framework is proposed that utilizes the Kolmogorov–Arnold Network (KAN) to capture complex features in the data, with the efficiency of decoupled knowledge distillation (DKD) training approach. A high-capacity KAN network is first trained to detect attacks performed on the test bed. This model then serves as a teacher to guide a much smaller multilayer perceptron (MLP) student model via DKD. The resulting DKD-MLP model contains only 2,522 and 1,622 parameters for WADI and SWaT datasets, which are significantly smaller than the number of parameters of the KAN teacher model. This is highly appropriate for deployment in resource-contained devices with limited computational resources. Despite its low size, the student model maintains a high performance. Our approach demonstrate the practicality of using KAN as a knowledge-rich teacher to train much smaller student models, without considerable drop in accuracy in intrusion detection frameworks. We have validated our approach on two publicly available datasets. We report F1-score improvements of 4.18\% on WADI and 3.07\% on SWaT when using the DKD-MLP model, compared to the bare student model. The implementation of this paper is available on our \href{https://github.com/BahMoh/DKD-KAN}{GitHub Repository}.
\end{abstract}


\section{Introduction}\label{sec:Introduction}
Cyber-physical systems (CPS) are composed of closely coupled physical components, sensing devices, actuators, and control units, which interact and coordinate via networked communication. These networks, however, are susceptible to cyber threats and unauthorized intrusions. CPS technologies have become prevalent across various domains such as healthcare, power grid systems, autonomous driving, transportation, and industrial manufacturing. As a modern class of systems, CPS blends computational intelligence with physical processes, enabling innovative modes of human interaction. By unifying computation, communication, and control, CPS significantly augments the capabilities of physical systems, ensuring their functionality, safety, and reliability in critical environments \cite{baheti2011cyber}.

Driven by continuous advancements in networking infrastructure, as well as the quick adoption of the internet of things (IoT) and industrial IoT (IIoT), in fields such as smart appliances, healthcare monitoring equipment, self-driving vehicles, and distributed sensor systems are becoming increasingly widespread \cite{abed2024modified}. As a result, the integration of CPS into various domains is expanding rapidly. In parallel, cyber-security has become a pressing issue affecting both conventional cyberspace and CPS-based environments. Within the framework of the fourth industrial revolution, where digital connectivity is deeply interwoven into many aspects of daily activities, ensuring a secure and robust cyberspace has gained critical importance \cite{muneer2024critical, mukherjee1994network, yick2008wireless}.

In IIoT settings, detection mechanisms are designed and incorporated to identify irregularities and anomaly patterns in sensor and actuator traffic data. As machine learning (ML) and deep learning (DL) technologies have advanced, there has been a transition from conventional detection techniques toward more adaptive, data-centric models \cite{alikhani2026learning, alikhani2025long}. ML-based approaches tend to be lightweight in terms of both computational demand and data requirements; however, they often encounter limitations when applied to unfamiliar or novel conditions. On the other hand, DL models benefit from greater representation capacity but require extensive datasets and considerable computational infrastructure. These requirements also introduce concerns about data confidentiality \cite{muneer2024critical}. In response, various research directions have been proposed. Unsupervised learning are developed to reduce the time and cost involved in manual data labeling \cite{li2024genos}, though its accuracy may not always meet expectations. To address privacy issues, federated learning has gained traction as a method that allows model training across decentralized data sources without needing to aggregate sensitive information \cite{muneer2024critical}. Moreover, a significant limitation in many traditional systems is the lack of transparency, which poses challenges in high-stakes industrial environments that require trust and accountability \cite{adadi2018peeking}.

To overcome issues such as scarcity of labeled data or uneven class distribution, numerous algorithms utilize oversampling methods like SMOTE, as demonstrated in \cite{lachure2024securing}, along with its various adaptations. Generative models based on GANs have also been investigated. Nevertheless, these approaches face several drawbacks. Firstly, their effectiveness declines sharply under conditions of extreme class imbalance, a prevalent challenge in many intrusion detection datasets. Secondly, because attack and normal data often share similar distributions, generating synthetic samples may obscure the boundaries between classes, leading to decreased detection performance. Additionally, GAN training tends to be unstable and computationally intensive. In this study, we avoid employing such augmentation or imbalance correction techniques. Instead, we implement a masking approach, randomly obscuring a portion of the data to introduce variability and improve the robustness of the model \cite{ghorbani2025using}.

Feature selection is also frequently applied in intrusion detection systems (IDSs) \cite{kravchik2021efficient}. However, a significant limitation of this method is that if an attack exploits a feature eliminated during the selection process, detection accuracy may suffer substantially. In this study, we exclude only features exhibiting no variation in the WADI and gas pipeline datasets and do not perform any further feature selection.

For classification tasks, pre-processing often presents a major bottleneck for fast, real-time inference, which can slow down the identification of cyber threats. Many current methods depend on sampling, where the algorithm must accumulate a full window of data before performing tasks such as averaging or extracting features using one-dimensional-CNNs, followed by classification. These windowing techniques often result in reduced detection performance \cite{de2020intrusion}.

Another frequently employed pre-processing method is principal component analysis (PCA), which lowers data dimensionality by projecting inputs onto orthogonal axes aligned with the directions of greatest variance. Although PCA can be beneficial, it risks disregarding features with low variance that are nonetheless important for precise classification. PCA may also diminish model interpretability since the transformed components do not have straightforward physical meanings.

In this work, we utilize an instance-level strategy, removing the dependency on windowing. Additionally, our approach bypasses complex pre-processing steps like PCA, employing only basic standard scaling. Initially, a powerful teacher model with many parameters is trained; then, through the DKD training procedure, a lightweight MLP student model with significantly fewer parameters is developed. This work’s key contributions can be summarized as:

\begin{itemize}
    \item DKD training methodology, resulting in significant model compression.
    \item A lightweight and extremely fast detection methodology, requiring minimal pre-processing.
    \item Retention of performance, even with extreme model compression.
    \item Validation of the method's effectiveness using two public datasets.
\end{itemize}

This paper is structured in the following manner: \autoref{sec_Related_Works} belongs to literature review. \autoref{sec_Datasets} introduces the datasets used in our experiments. \autoref{sec_Method} details the proposed approach, including the Kolmogorov–Arnold theorem and KAN architecture, data scaling, knowledge distillation (KD), DKD, and the simulation setup. The findings of our experiments are detailed in \autoref{sec_Results}. And , \autoref{sec_Conclusion} offers the conclusion.

\section{Related Works}\label{sec_Related_Works}
\cite{gao2019adaptive} proposes an adaptive ensemble learning model for intrusion detection. A multi-tree algorithm is developed by adjusting training data proportions, using multiple decision trees. Several classifiers, including decision tree, random forest, kNN, and deep neural network (DNN), are combined through an adaptive voting mechanism. 
\cite{rajadurai2022stacked} suggests a stacked ensemble classifier, addressing the limitations of single algorithms in handling large-scale network data. The method is compared against several popular ML models, including artificial neural network (ANN), random forest, and support vector machine (SVM). The main limitations of the above methods are twofold: 1) the reliance on feature selection, and 2) low generalizability of the tree-based approaches.

The application of DL techniques, particularly CNN and LSTM architectures, has been extensively explored in the existing literature. In \cite{zhang2025network}, the authors propose a hybrid CNN-LSTM model aimed at improving the representation of temporal patterns within IIoT networks. Despite its potential, the study acknowledges persisting challenges such as data imbalance and privacy concerns, and recommends the integration of federated learning in future work.

A recurrent neural network (RNN) is proposed in \cite{yin2017deep}, under binary class and multiclass scenarios. And the performance of the model is evaluated again traditional ML approaches, such as SVM, random forest and ANN. This work show cases the effectiveness of DL methods.
To address the changing and evolving attack scenarios, \cite{vinayakumar2019deep} proposes a DL-based cyber-attack detection. To acquire a robust evaluation of the method, the network is tested on various datasets to simulate the varying attacks. The problem with this approach is long training time, that requires 1000 epochs.

In \cite{shone2018deep}, the problem of high reliance of intrusion detection on human input and declining detection accuracy, using stacked nonsymmetric deep AE (NDAE) in an unsupervised manner. The shortcoming of this approach is that it was evaluated on only one dataset. 
\cite{khan2019novel} also investigates new threats that could not be identified by existing IDS. To do so, a two-stage DL method is proposed wite utilizes stacked autoencoder (AE). The first stage is responsible for detecting normal and abnormal traffic data with a probability score. The second stage, aids the first stage with additional features. The training is performed using large volumes of unlabeled data.

In \cite{qazi2023hdlnids} a hybrid convolutional RNN (CRNN) is applied for intrusion detection. In this work, CNN is used to detect local features in the data, and the RNN is used for fusing different features created by the CNN network. 
\cite{imrana2021bidirectional} proposed a bidirectional LSTM (BiLSTM) network to solve the issue of high false alarm rate in the traditional IDS.

Meanwhile, in \cite{feng2021time}, the authors adopt a state-space modeling approach, using neural networks to identify the system dynamics of a physical process. Anomaly detection is then achieved through Bayesian filtering by assessing discrepancies between the predicted and actual system behavior.

\cite{wang2025feco} combines federated learning with contrastive training for IoT devices intrusion detection. The use of federated learning addresses the privacy concerns, since the training is performer locally on the user's device. And also, a two-stage feature selection to reduce overfitting and inference time. However, the use of feature selection, can hinder the ability of the network to detect intrusions if the attacker targets the discarded features.
In \cite{lopez2022contrastive}, with the use of shallow neural network and contrastive training, tries to address the treat of evolving attacks. Here both labels and features are moved to the same embedding space, and class labels act as a prototype for the features. The contrastive learning aligns the projected features with the class prototypes. The model is tested on the zero-shot learning constraints as well.

To address the problem of generalizability to unseen distributions, \cite{golchin2024sscl} proposes a semi-supervised contrastive learning-based IDS. This framework is pretrained on the normal data, and fine-tuned on twenty data samples. 
\cite{zhang2024aoc} discusses the problem of offline training of IDS, and states that in a real-world scenario, the attacker's behavior undergoes changes and this can limit the IDS performance, so adaptive IDSs must be proposed. This framework employs an AE with cluster repelling loss function to train the AE with the incoming data, in an online manner.
Meanwhile,~\cite{alikhani2025contrastive} employs a combination of KAN and contrastive learning for IoT intrusion detection, specifically addressing the challenge of sparse labeled data on three public datasets. In this work, KAN outperforms MLPs and other contrastive methods on multiple datasets using minimal labeled data, offering high accuracy, robustness, and interpretability. The problem with these DL-related methods is that they have large number of parameters, which limits the applicability of these methods on resource-constrained devices.

\cite{caville2022anomal} uses graph neural networks (GNNs) to simulate the flow of network traffic which is like a graph. This method addresses the requirement for labeled data and also disregards the traditional GNN assumptions for node features that could degrade the performance of intrusion detection. 
The work in \cite{nguyen2023ts} addresses the limitations of the traditional ML and DL-based IDSs that extract the features independently and ignores the interactions of the network elements with each other. To counter this issue, the authors propose a self-supervised traffic-aware IDS to learn the relationships between the network nodes.
\cite{xu2024applying} proposes a GNN for IDS, that utilizes a graph-attention encoder to obtain edge features in an unsupervised manner, and a graph contrastive learning is applied to create negative and positive samples from subgraphs. Using these positive and negative samples the entire graph is trained. 

\cite{ashfaq2017fuzziness} uses fuzzy logic and DL to detect intrusions. In this work, with the use of a single layer neural network, and semi-supervised leaning on large amount of unlabeled data, the neural network is trained to output membership vectors for sample fuzziness categorization, low, mid and high.
A multi-level semi-supervised learning method is proposed in \cite{yao2018msml} to address the imbalance of the dataset and the domain deference between the train and test sets.
In \cite{madhuri2024new} a neural network with hierarchical k-means method is proposed in conjunction with semi-supervised learning. The Grasshopper optimization is used to optimize the parameters of the neural network.
A semi-supervised IDS is proposed in \cite{abdel2021semi}, which utilizes a multiscale residual temporal convolutional network with an attention mechanism that focuses the intrusion detection on the impropriate features. Also a hierarchical module is introduced to give weight to the sequential properties of the network data.

Given the above mention limitations of the DL methods, which is the high number of parameters of the network which hinders the applicability of the designed IDS on low-end systems, KD-related methods are proposed to tackle the problem of large models in DL. 
\cite{wang2022lightweight} utilizes a CNN along with with triplet loss function and k-fold cross-validation to train a lightweight IDS. 
In \cite{shen2024effective}, the problem of heterogeneity between different clients is tackled using a federated learning ensemble KD (FLEKD), avoiding sharing the training data, in a centralized server. This approach, proves to be better than the conventional model fusion methods.
The work in \cite{yang2023lightweight} proposes batch-wise self-knowledge distillation (SKD) for intrusion detection to address the performance degradation that is caused by smaller number of model parameters.

\cite{wang2024lightweight} combines CNN and KD to address data sparsification caused in high dimensional datasets. This framework employs the Fourier transform to convert signals from the time domain to the frequency domain, to aid feature continuity. The training is performed with adaptive temperature KD. The teacher is a CNN model with eight layers, however with effective training the student model, accurately detects intrusions, with only one a one layer CNN network, significantly compressing the model. In \cite{wang2023cybersecurity}, the CSNT knowledge graph completion model is proposed, utilizing BiLSTM for capturing entity-relationship interactions, and combining neural networks with tensor decomposition. To address the problem of catastrophic interference, forgetting previously learned knowledge, it utilizes SKD.

\cite{de2024vincent} introduces VINCENT, a DL method using vision transformers (ViTs) for network intrusion detection. VINCENT encodes features into color images, that allows the transformer to extract explainable class signatures through the self-attention mechanism. Using KD and the trained transformer, a lightweight student model is trained for the task.
\cite{xie2025dtkd} introduces dual-teacher KD IDS (DTKD-IDS), for intrusion detection in IIoT networks. The model employs prototype distillation by extracting valuable knowledge from two teacher networks based on data prototype vectors. 
\cite{zou2024cyber} focuses on cyber-attack detection and intervention. To address limitations like non-IID network traffic and blurred boundaries between normal and attack samples, the method incorporates KD and prototype aggregation into federated learning to improve detection accuracy and computational efficiency. 

\section{Datasets} \label{sec_Datasets}
This section explores the different ICS datasets used in this research. Several publicly available datasets are commonly cited in the literature, including SWaT \cite{mathur2016swat}, WADI \cite{ahmed2017wadi}, Gas Pipeline \cite{morris2011control}, \cite{morris2014industrial}, BATADAL \cite{taormina2018battle}, among others. Additionally, there are IoT-related datasets such as UNSW‐NB15 and ToN‐IoT, although these are outside the scope of this discussion. Datasets generally fall into two main types: simulated datasets, including BATADAL, and real-world testbed data, e.g. SWaT, WADI, and the Gas Pipeline dataset. One of the key drawbacks of the BATADAL dataset is that it is simulated, with a sampling time of one hour, and more critically, it is partially labeled, meaning that there are errors in the annotations. Therefore, this study focuses on the testbed datasets. We validate our proposed method using the SWaT and WADI datasets, which primarily consist of physical measurements from sensors and actuators, although network features may also be included. A detailed discussion of each dataset follows \cite{alikhani2025contrastive}.

\subsection{SWaT Dataset}
The SWaT testbed was introduced by the iTrust Center at the SUTD to evaluate the effects of cyber-attacks on industrial systems, as well as detection methods, defense mechanisms, and the cascading effects of attacks on other ICS components. The system consists of six distinct sub-processes, which are controlled by its own PLCs. These PLCs communicate through wired or wireless channels and are connected to a SCADA unit. The SWaT dataset includes both network data and physical data, such as sensor and actuator readings. However, for this study, we focus solely on the physical data gathered from a small testbed. The dataset includes several versions, spanning 2015, 2017, 2019, and 2020, with the 2015 version being the most commonly studied. To ensure better comparison across methods, we used the 2015 version. The dataset is composed of two files: one containing normal data, recorded over seven consecutive days under normal plant conditions, and the other containing attack data, recorded over four days under various attack scenarios. This dataset has been primarily studied under a binary classification assumption (normal vs. attack). The normal file is generally used for unsupervised and one-class learning approaches, where only normal data is fed to the network during training. In our study, we focus on the attack file, which contains normal and attack instances. The dataset columns are consist of 51 sensor and actuator readings, along with timestamps sampled at approximately one-second intervals. The target labels are included as a column in the dataset files \cite{mathur2016swat}. The SWaT dataset consists of 87.86\% normal samples (395,298) and 12.14\% attack samples (54,621) \cite{ghorbani2025using}.

\subsection{WADI Dataset}
The WADI dataset, developed by the SUTD, is a small-scale water distribution system designed to support research on secure water distribution systems and the cascading effects of cyber-attacks on ICS, including other systems like power generation and distribution, as well as the SWaT dataset. The WADI testbed consists of three main phases, each controlled by a PLC: the primary grid (P1), which can be derived from the output of the SWaT testbed; the secondary grid (P2); and finally the return water grid (P3). The PLCs are communicate with sensors and actuators via a wireless network, while another wireless network links the PLCs to the SCADA system. The WADI dataset also has different versions, with the 2017 version being the most studied in the literature. It includes two files: A normal file, which the testbed is operated during 14 consecutive days, under normal condition, and an attack file with two days of various attack scenarios. Similar to SWaT dataset, normal and attack instances are present in the attack file. For our study, we use the attack file in a binary classification format, distinguishing between attack and normal conditions. In WADI dataset the normal-attack imbalance ratio is higher than the SWaT dataset. Although attack labels are not directly included in the dataset files, a PDF document is provided that contains a timetable recording the exact times when attacks occur. However, this timetable contains errors, as some attack occurrences are incorrectly recorded. The original WADI dataset has 130 features, including time stamps, dates, and indexes. After removing these features, 127 features remain, four of which contain only NaN values, so these columns are dropped. Ultimately, the dataset consists of 123 sensor and actuator features \cite{ahmed2017wadi}. The WADI dataset consists of 94.24\% normal samples (162,853) and 5.76\% attack samples (9,948) \cite{ghorbani2025using}.

In both the WADI and SWaT datasets, there is a period of time allocated for the system to stabilize after each attack, before another attack is carried out on the system. \autoref{tab_swat_dataset} and \autoref{tab_wadi_dataset} provide a concise overview of the characteristics of the two datasets.
\begin{table}[htbp]
\caption{SWaT Dataset Properties and Train-Test Sizes}
\label{tab_swat_dataset}
\centering
\begin{tabular}{@{}lllll@{}}
\toprule
\textbf{Dataset} & \textbf{Total Features (Used)} & \textbf{Train Size} & \textbf{Test Size} \\
\midrule
SWaT & 52 (51) & 359,935 & 89,984 \\
\bottomrule
\end{tabular}
\end{table}

\begin{table}[htbp]
\caption{WADI Dataset Properties and Train-Test Sizes}
\label{tab_wadi_dataset}
\centering
\begin{tabular}{@{}lllll@{}}
\toprule
\textbf{Dataset} & \textbf{Total Features (Used)} & \textbf{Train Size} & \textbf{Test Size} \\
\midrule
WADI & 130 (123) & 138,240 & 34,561 \\
\bottomrule
\end{tabular}
\end{table}

\section{Methodology}\label{sec_Method}
In this section, first KAN is discussed in \autoref{subsec_KAN}. In \autoref{subsec_preprocessing}, we discuss the preprocessing steps associated with the proposed method. Section \autoref{subsec_KD} outlines the general KD approach, and in \autoref{subsec_DKD}, we explain the proposed DKD.

\subsection{KAN}\label{subsec_KAN}
KAN is a type of fully connected network proposed in \cite{liu2024kan}, designed to leverage the Kolmogorov–Arnold representation theorem (KART) by replacing the learnable weights in the network with one-dimensional activation functions whose outputs are aggregated at the nodes. In traditional fully connected networks, such as MLPs, learnable linear weights are placed on the edges between nodes, and nonlinearity is introduced after the aggregated outputs pass through a nonlinear activation function at each node. \cite{funahashi1989approximate} is one of the earliest works to establish a connection between KART and neural networks. This work, with the help of KART, proves that any function can be approximated using a neural network with at least two layers and sigmoid activation functions. 

In a relatively short time, KAN has attracted significant attention from researchers in the field of artificial intelligence across various domains. Examples include medical applications, such as medical image segmentation \cite{li2025u}, engineering applications, such as remote sensing \cite{cheon2024kolmogorov}, and nonlinear function approximation, where a combination of Chebyshev polynomials and KAN is employed \cite{ss2024chebyshev}. One of KAN’s notable advantages is the ability to visualize spline functions, thereby enhancing the interpretability of the model. Since the use of spline functions enables KAN, in a deterministic manner, to assess the importance of each feature and to characterize the properties of each learned function, it offers a transparent and interpretable framework for understanding model behavior.

One of the areas that KAN has been extensively studied, are the fields of time series classification and anomaly detection that are tightly related to intrusion detection and cyber-security. So we first examine the applications of KAN in these fields. In \cite{alikhani2025kan}, a KAN is proposed for human activity classification using wearable sensor data and smartphone IMU sensors. In this work, only three-axis accelerometer data are used for classification, the authors prove that the proposed framework is capable of competing with state of the art approaches. \cite{yu2025kanids} integrates temporal graphs with KANs to capture system behavior for detecting stealthy, long-term cyber-attacks. While MLPs could be used for this purpose, KAN is adopted due to its superior modeling capabilities. 

In \cite{dutta2025kol}, authors propose a malware classification, combining KANs and GANs to tackle data imbalance. \cite{saber2025kolmogorov} uses KAN for cyber-attack detection in electric vehicle charging infrastructure. By analyzing only the power consumption measurements, this approach can distinguish between normal and abnormal conditions. A hybrid intrusion detection algorithm is proposed in \cite{saritha2025novel} that utilizes GRUs for feature extraction and KANs for final classification. \cite{zhao2025lightweight} proposes a convolutional KAN (C-KAN) for autonomous vehicle cyber-attack detection. First, using GAN synthetic attack data are generated, then using a lightweight C-KAN the detection is performed. In \cite{sriraghavendra2025knowledge}, the authors propose an intrusion detection method for sensor networks. In this approach, a knowledge graph is designed by experts, features are extracted using PCA and DL, and KAN decomposes complex relationships between features.

KART states that any $T$-dimensional continuous function $f(X)$, where $X = \{ X_i \}_{i=1}^T$, can be expressed with the sum of finite number of one-dimensional continuous functions. This function can be represented with one-dimensional functions \( \Phi_q \) and \( \phi_{q,p} \) as follows:

\begin{equation}
f(X)=KAN(X) = \sum^{2T+1}_{q=1}\Phi_q \bigl ( \sum^T_{p=1}\phi_{q,p}(X_p) \bigr )
\label{eq:KA_theorem}
\end{equation}
here each \( \Phi_q \) and \( \phi_{q,p} \) is a one-dimensional function. This summation empowers the theorem to approximate any multivariate function, with finite number of one-dimensional functions, which is $T \times (2T+1)$. One of the advantages of KART compared to other function approximation methods is the finite number of elements required to fully satisfy the function approximation, whereas, Fourier series or Taylor expansion requires infinite number of elements, smaller number of elements will result in lower accuracy. These functions, \( \Phi_q \) and \( \phi_{q,p} \), are learnable activations that will be trained during training. Unlike MLPs, where activation functions are applied to the output of neurons, in KAN the activations are applied to the links between the neurons and the activations are aggregated in the neurons.

Taking a closer look at the \autoref{eq:KA_theorem}, it is evident that there is a large similarity between this equation and the standard MLP formulation, which is as follows:

\begin{equation}
f(X)=MLP(X) = \sum_{q} v_{q} \, \sigma\!\left( \sum_{p} w_{pq} X_{p} + b_{q} \right)
\label{eq_MLP}
\end{equation}
where, $\sigma(\cdot)$ is the nonlinear function, $w_{pq}$ are the trainable weights. It is evident that \autoref{eq:KA_theorem} and \autoref{eq_MLP} share similarities. Both formulas are superposition of the one-dimensional nonlinearities applied to the combinations of inputs. So the KART theorem can be used as an alternative to MLP neural networks.
Keep in mind that there are differences as well. For example, for the MLP equation, the inner summation is composed of linear mappings, and the nonlinearity is applied in the outer summation. This makes KAN much more powerful and effective than MLPs.

\begin{figure}
    \centering
    \includegraphics[width=0.5\linewidth]{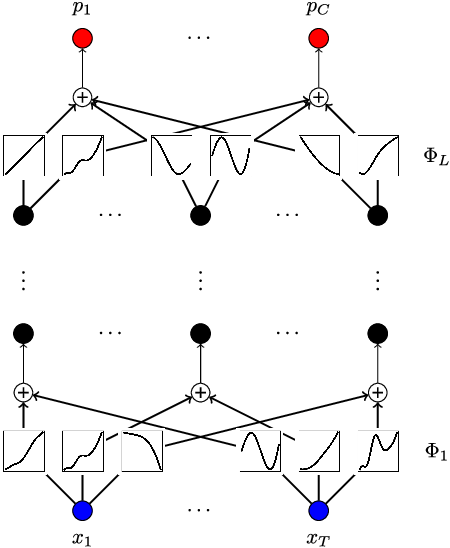}
    \caption{A \( L \)-layer stacked KAN network with \( T \) inputs and \( C \) output classes, used for classification tasks \cite{baravsin2024exploring}}
    \label{fig_KAN_architecture}
\end{figure}

\autoref{fig_KAN_architecture} illustrates an $T$-variate KAN with $L$ layers, which outputs $C$-dimensional predictions. This network will be used for a $C$-class classification task, which in our case for a binary classification task $C=2$. Here, $KAN(\cdot)$ denotes the network, $\{X^i,Y^i \}$ denotes the data, where the label $Y^i$ belongs to the classes $\{1,\dots,C\}$. The class representations will be generated as $p^i=KAN(X^i)$, where $p^i=\{1, \dots,C\}$, but for simplicity and uniformity, we denote the class representations with $y^i$ to be in the same direction as the mentioned functionality of KART, which is function approximation.

For the data point $X^i$, the forward pass through the specified KAN classifier, can be defined as a cascade of $L$ different nonlinear transformations, $\Phi_{l}$ where $l \in \{1, \dots,L\}$, as follows:
\begin{equation}
y = KAN(X^i) = (\Phi_L \circ \Phi_{L-1} \circ \dots \circ \Phi_{1})X^i
\label{eq:kan_in_matrix_form}
\end{equation}
here, $\circ$ denotes function composition. In contrast, MLPs perform  the forward pass as a sequence of linear transformations followed by nonlinear activations:
\begin{equation}
y = MLP(X) = (W_L \circ \sigma \circ W_{L-1} \circ \sigma \circ \dots \circ W_{1})X
\label{eq_mlp_in_matrix_form}
\end{equation}

Since MLPs need to exhibit nonlinearity, nonlinear activation functions $\sigma$ are introduced in the architecture. During the training, KAN learns the activation functions, $\Phi$ and $\phi$, however, MLP learns the linear weights $W$.

For these one-dimensional learnable activation functions, KAN incorporates a combination of spline basis functions with sigmoid linear unit (SiLU) activation functions, which are as follows:
\begin{equation}\label{eq:silu_spline}
\phi(X) = \omega_bSiLU(X)+\omega_sSpline(X)
\end{equation}
the SiLU function is defined as follows:
\begin{equation}\label{eq:silu_activation}
SiLU(X) = \frac{X}{1 + e^{-X}}
\end{equation}

The SiLU activation combines the favorable characteristics of both the sigmoid and the rectified linear unit (ReLU). From the sigmoid, it inherits smooth differentiability, unlike ReLU which is non-differentiable at zero. Moreover, SiLU allows small negative outputs for slightly negative inputs, reducing the risk of dead neurons. Unlike sigmoid, it does not suffer from saturation, and similar to ReLU, it grows approximately linearly for large positive inputs.

The B-splines are a family of piecewise polynomial functions, which form a basis set for representing smooth curves. B-splines of order $K$ create a polynomial of the same order in the interval. And they perform the computations in a grid of size $G$. 
\begin{equation}\label{eq:spline_basis_function}
Spline(X) = \sum ^{G+K}_{i=1} c_iB_i(X)
\end{equation}
$c_i$ is a trainable coefficient known as control point, and $B_i$ is the spline basis functions. The main parameters, contributing to the performance of the KAN are the grid size denoted by $G$ and the spline order denoted by $K$, where it determines the smoothness of the curve. In classification task, higher grid sizes and lower spline orders will result in better performance.

The parameter count of KAN, depends on four factors, which are: 1) input size, 2) output size, 3) grid size, and 4) spline order. The formula for parameter count is as follows:

\begin{equation}\label{eq:KAN_parameters}
Parameters = (d_{in} \times d_{out})(G+K+3) + d_{out}
\end{equation}
here, $G$ and $K$ were discussed before. $d_{in}$ and $d_{out}$ denote the input and output dimensions. From this formula, it is evident that the number of parameters for KANs is much higher than that of MLP networks.

\subsection{Data scaling}\label{subsec_preprocessing}
Using the mean ($\mu$) and standard deviation ($\sigma$) of each feature we perform the standardization of the data. KAN has the distinct advantage of being able to classify directly without requiring any additional data preprocessing, leading to faster inference times \cite{alikhani2025contrastive}.

The standardization is carried out using the following procedure:
\begin{equation}
X_{\text{scaled}} = \frac{X - \mu}{\sigma}
\label{eq_standard_scaler}
\end{equation}

We also experimented with min-max scaling, but found that standardization yielded better results.

\subsection{Knowledge distillation}\label{subsec_KD}
KD is a popular method for compressing DL models in which a low-parameter student model, is trained to mimic the output of a powerful, high-parameter teacher model. Through this process, the student model achieves results similar to the teacher’s, while reducing the number of parameters and computational resources required for the model deployment \cite{petrosian2024dkdl, gou2021knowledge}.

In real-time ML and DL applications, models with fewer parameters are favored due to their reduced computational costs and often improved generalization. In this approach, first, a teacher model, based on KAN is trained, and then the teacher model provides supervision to a lightweight student model built using an MLP, which maintains high performance while having considerably fewer parameters.

\subsection{Decoupled knowledge distillation}\label{subsec_DKD}
The DKD technique is an enhanced version of KD that divides the distillation process into separate components. Rather than combining the knowledge from both the correct class and incorrect classes, DKD separates them into distinct loss components, each with its own weight. In DKD, the distillation loss is split into two independent parts: one for target and one for the non-target classes. Each component is weighted using a specific hyperparameter. This separation affords the student to learn subtle distinctions between the target and non-target classes, which may be lost when using a single, coupled KL loss function.

To begin, a binary probability vector $b=[p_t, p_{-t}] \in \mathbb{R}^{1 \times 2}$ is defined on the output of the model, where $p_t$ is the target class probability, and $p_{-t}$ is the total probability for all non-target classes (excluding the target class). These probabilities are defined as:
\begin{align}
&p_t =\frac{\exp \left(l_t\right)}{\sum_{j=1}^N \exp \left(l_j\right)}\label{eq_target_prob} \\
&p_{-t} =\frac{\sum_{d=1, d \neq t}^N \exp \left(l_d\right)}{\sum_{j=1}^N \exp \left(l_j\right)} \label{eq_non_target_prob}
\end{align}
$\tilde{p}_i$ is the probability corresponding to the non-target categories, and it is as follows:
\begin{align}
\tilde{p}_i & =\frac{p_i}{p_{-t}}=\frac{\exp \left(l_i\right)}{\sum_{j=1, j \neq t}^N \exp \left(l_j\right)} 
\label{eq_prob_non_target}
\end{align}
where $p_i$ is the normal softmax computed for $i^{\text{th}}$ class, while $p_{-t}$ is the total probability for all non-target classes, calculated as $p_{-t}=1-p_t$. The mathematical form of the softmax function is:
\begin{equation}
    p_i =\frac{\exp \left(l_i\right)}{\sum_{j=1}^N \exp \left(l_j\right)}
    \label{eq_softmax}
\end{equation}
KD employs the Kullback–Leibler (KL) loss function. Initially, we present the standard form of the KL divergence, and then reformulate it using the binary probability vector $b$ along with the non-target class distribution $\tilde{p}$: 
\begin{equation}
\mathcal{L}_\mathrm{KD}={\mathrm{KL}}\left(p^T \| p^S\right)=p_t^T \log \frac{p_t^T}{p_t^S}+\sum_{i=1, i \neq t}^N p_i^T \log \frac{p_i^T}{p_i^S}
\label{eq_KLD_loss}
\end{equation}
here, $S$ and $T$ represent the student and teacher probabilities, which are the MLP and KAN models. The function ${\mathrm{KL}}$ denotes the Kullback–Leibler divergence, which measures the difference between the smoothed probability distributions predicted by the student and those predicted by the teacher.
Expanding \autoref{eq_KLD_loss} using the definitions from \autoref{eq_prob_non_target} and \autoref{eq_softmax} yields the following expression:

\begin{equation}
    \mathcal{L}_\mathrm{KD}=p_t^T \log \left(\frac{p_t^T}{p_t^S}\right)+p_{-t}^T \sum_{i=1, i \neq t}^N \tilde{p}_i^T\left(\log \left(\frac{\tilde{p}_i^T}{\tilde{p}_i^S}\right)+\log \left(\frac{p_{-t}^T}{p_{-t}^S}\right)\right)
\end{equation}

This can be simplified into:
\begin{equation}
\mathcal{L}_{\mathrm{KD}}=p_t^T \log \left(\frac{p_t^T}{p_t^S}\right)+p_{-t}^{T} \log \left(\frac{p_{- t}^T}{p_{- t}^S}\right)+p_{-t}^T \sum_{i=1, i \neq t}^N \tilde{p}_i^T \log \left( \frac{\tilde{p}_i^T}{\tilde{p}_i^S}\right)
\label{eq_expanded_KLD_loss}
\end{equation}

\autoref{eq_expanded_KLD_loss} can be rewritten as:
\begin{equation}
\mathcal{L}_{\mathrm{KD}} = {\mathrm{KL}}\left(\mathbf{b}^T \| \mathbf{b}^S\right)+\left(1-p_t^T\right) \cdot {\mathrm{KL}}\left(\tilde{p}^T \| \tilde{p}^S\right) 
\end{equation}
here, ${\mathrm{KL}}\left(\mathbf{b}^T | \mathbf{b}^S\right)$ quantifies the divergence between the teacher and student probabilities for the target and non-target classes. This can be decoupled into two different elements: the target class KD ($\mathrm{TCKD}$) and non-target class KD ($\mathrm{NCKD}$). With these definitions, the distillation loss can be rewritten as:
\begin{equation}
\mathrm{KD} =\mathrm{TCKD}+\left(1-p_t^T\right) \cdot \mathrm{NCKD}
\label{eq_final_KD}
\end{equation}
where $\mathrm{TCKD} := \mathrm{KL}(\mathbf{b}^T \| \mathbf{b}^S)$ and $\mathrm{NCKD} := \mathrm{KL}(\tilde{p}^T \| \tilde{p}^S)$. As seen in \autoref{eq_final_KD}, in classical KD the weights are coupled. For well-predicted samples the $(1-p^T_t)$ becomes small, that decreases the effect of $\mathrm{NCKD}$ and the ability of the framework for changing the weights of each term in order to more effectively balance the importance of each loss. Therefore, to address the challenges posed by classical KD, decoupled KD introduces independent hyper-parameters $\alpha$ and $\beta$ to control $\mathrm{TCKD}$ and $\mathrm{NCKD}$ separately, as follows:
\begin{equation}
\mathcal{L}_{\mathrm{DKD}} =\alpha \cdot \mathrm{TCKD}+\beta \cdot \mathrm{NCKD}
\end{equation}
where $\alpha$ is the weight of the target class and $\beta$ is the weight of the non-target class. By adopting this, the knowledge gained by the teacher model, transfers better to the student model. By decoupling the contributions of the target and non-target classes and introducing tunable parameters $\alpha$ and $\beta$, DKD provides more control over the distillation process, making it more flexible and potentially more effective, particularly in cases where the student struggles to capture the subtle differences between classes \cite{petrosian2024dkdl}.

The overall training loss is a weighted combination of the standard cross-entropy loss using ground-truth labels and the DKD loss derived from the teacher’s soft predictions:
\begin{equation}
\mathcal{L}_{\text{total}} = (1 - \lambda) \cdot \mathcal{L}_{\text{hard}} + \lambda \cdot \mathcal{L}_{\text{DKD}}
\end{equation}
and to prevent the student model from overfitting to the teacher too early, a linear warm-up, denoted by $w_{\text{warmup}}$, is applied to gradually increase the contribution of the DKD loss during the early training epochs: 
\begin{equation}
\mathcal{L}_{\text{total}} = (1 - \lambda) \cdot \mathcal{L}_{\text{hard}} + \lambda \cdot \min\left(\frac{\text{epoch}}{w_{\text{warmup}}}, 1\right) \cdot \mathcal{L}_{\text{DKD}}
\end{equation}

\subsection{Simulation Setup} \label{sec_setup}
In this work, the parameters of the KAN are selected based on the guidelines provided in \cite{dong2024kolmogorov}, which suggest that networks configured with high grid sizes and low spline orders, despite exhibiting higher Lipschitz constants, tend to achieve better performance during training. Also the experiments in \cite{alikhani2025contrastive} confirms this behavior. The spline order and grid size are set to 1 and 50 for the WADI dataset, and to 3 and 50 for the SWaT dataset, respectively. All experiments are conducted in the Kaggle environment.

The training parameters are configured as follows: for the SWaT dataset, we use a warm-up constant $w_{\text{warmup}} = 80$, a regularization coefficient $\lambda = 0.1$, $\alpha = 5$, and $\beta = 1$; for the WADI dataset, we use $w_{\text{warmup}} = 5$, $\lambda = 0.2$, $\alpha = 5$, and $\beta = 1$. The parameter tuning for the SWaT dataset is more difficult since smaller warm-up parameter can lead to training instability. Furthermore, to smooth the logits produced by both the teacher and student models, temperature scaling is applied.

\begin{table*}[]
\caption{SWaT comparative study table}
\label{tab_swat_comparison}
\begin{tabular}{lllllll}
\toprule
\textbf{Reference} & \textbf{Methodology} & \textbf{Acc (\%)} & \textbf{Prec (\%)} & \textbf{Rec (\%)}  & \textbf{F1 (\%)} & \textbf{Number of Params.}\\ \midrule
\cite{audibert2020usad} & USAD        & -                          & 98.70          & 74.02           & 84.60 & -\\
\cite{zong2018deep}   & DAGMM         & -                          & 90.00          & 80.72           & 85.38 & -\\
\cite{park2018multimodal}  & LSTM+VAE & -                          & 98.39          & 77.01           & 86.40 & -\\
\cite{su2019robust} & OminiAnimaly    & -                          & 99.01          & 77.06           & 86.67 & -\\
\cite{zhang2019deep} & MSCRED         & -                          & 98.43          & 77.69           & 86.84 & -\\ 
\cite{li2019mad} & MAD-GAN            & -                          & 98.72          & 77.60           & 86.90 & -\\ 
\cite{deng2021graph} & GDN            & -                          & 98.85          & 91.42           & 93.59 & -\\ 
\cite{chen2021learning} & GTA         & -                          & 94.83	        & 88.10           & 91.34 & -\\
\cite{shen2020timeseries} & GTA       & -                          & 98.08          & 79.94           & 88.09 & -\\ 
\cite{ghorbani2025using}  & KAN       & \textbf{99.97}             &\textbf{99.89}  & \textbf{99.90}  & \textbf{99.89} & 22,260\\
\midrule
Proposed Method     & Teacher MLP model &  \textbf{99.94}          & \textbf{99.42} & \textbf{99.64}  & \textbf{99.53}& 87,450 \\ 
Proposed Method     & Student KAN model &  99.05            & 98.59          & 84.96           & 91.27& \textbf{1,622} \\
Proposed Method     & DKD-MLP           &  99.49            & 98.90          & 92.24           & 95.45& \textbf{1,622}\\ \bottomrule
\end{tabular}
\end{table*}

\begin{table*}[]
\caption{WADI comparative study table}
\label{tab_wadi_comparison}
\begin{tabular}{lllllll}
\toprule
\textbf{Reference} & \textbf{Methodology} & \textbf{Acc (\%)} & \textbf{Prec (\%)} & \textbf{Rec (\%)}  & \textbf{F1 (\%)} & \textbf{Number of Params.}\\ \midrule
\cite{audibert2020usad}    & USAD              & -                 & 64.51           &  32.20         & 42.96  & -      \\
\cite{zong2018deep}        & DAGMM             & -                 & 22.28           &  19.76         & 20.94  & -      \\
\cite{park2018multimodal}  & LSTM+VAE          & -                 & 46.32           &  32.20         & 37.99  & -      \\ 
\cite{su2019robust}        & OminiAnimaly      & -                 & 26.52           &  97.99         & 41.74  & -      \\ 
\cite{zhang2019deep}       & MSCRED            & -                 & 30.26           & 40.35		  & 34.58  & -      \\ 
\cite{li2019mad}           & MAD-GAN           & -                 & 41.44	         & 33.92          & 37.30  & -      \\ 
\cite{deng2021graph}       & GDN               & -                 & 85.62           & 85.41          & 85.52  & -      \\ 
\cite{chen2021learning}    & GTA               & -                 & 83.91           & 83.61          & 83.76  & -      \\ 
\cite{shen2020timeseries}  & GTA               & -                 & 42.12           & 63.34          & 50.59  & -       \\ 
\cite{ghorbani2025using}   & KAN               & \textbf{99.98}    & \textbf{99.85}  & \textbf{99.90}  & \textbf{99.87}   & 17,500 \\ \midrule
Proposed Method            & Teacher KAN model & \textbf{99.88}    & 98.84           & \textbf{99.09} & \textbf{98.96} & 198,750\\
Proposed Method            & Student MLP model & 99.48             & 97.21           & 93.62          & 95.38 & \textbf{2,522}\\
Proposed Method            & DKD-MLP           & 99.82             & \textbf{98.88}  & 98.02          & 98.45 & \textbf{2,522} \\ \bottomrule
\end{tabular}
\end{table*}

\begin{figure}[p]
    \centering
    \begin{tabular}{cc}
        \includegraphics[width=0.43\textwidth]{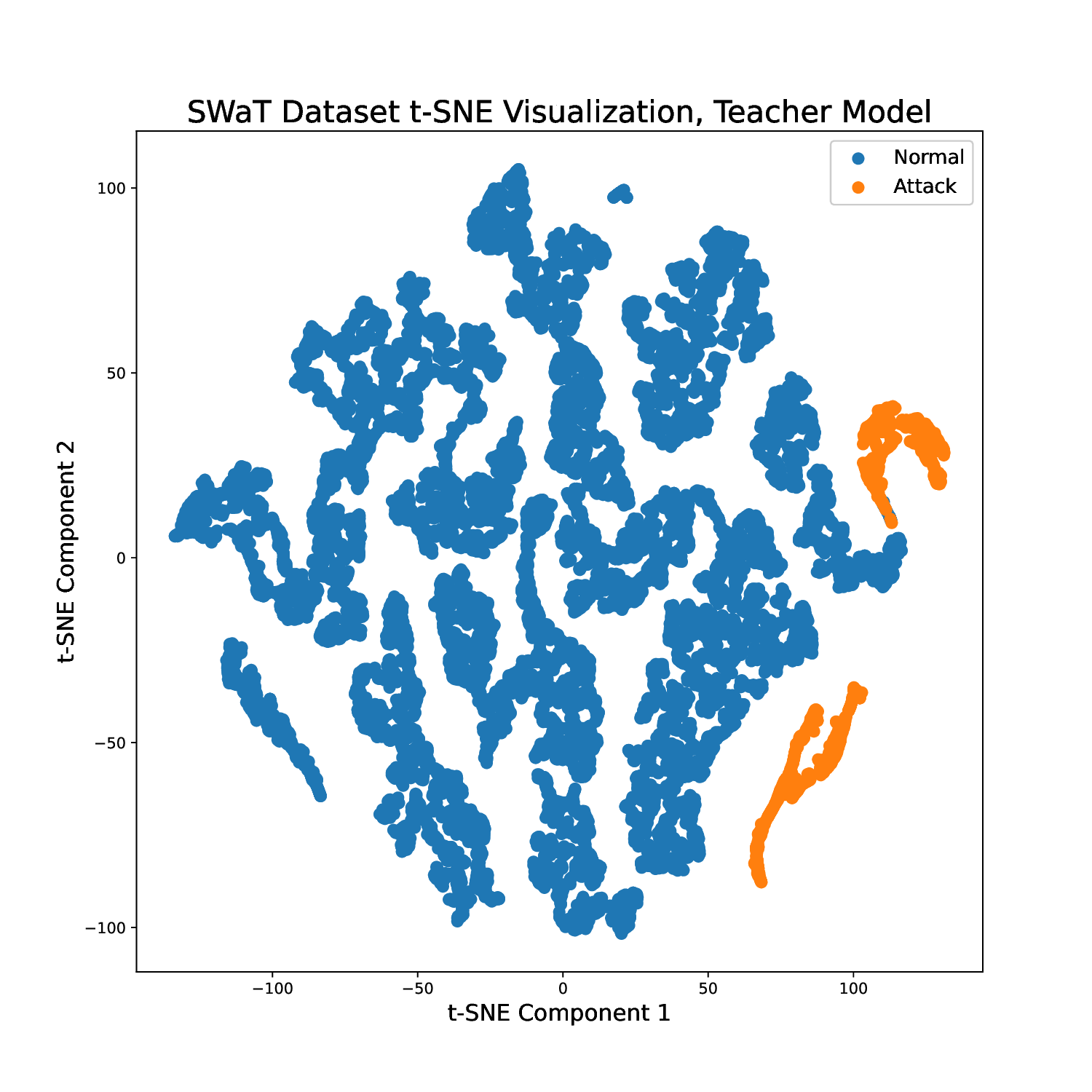} &
        \includegraphics[width=0.43\textwidth]{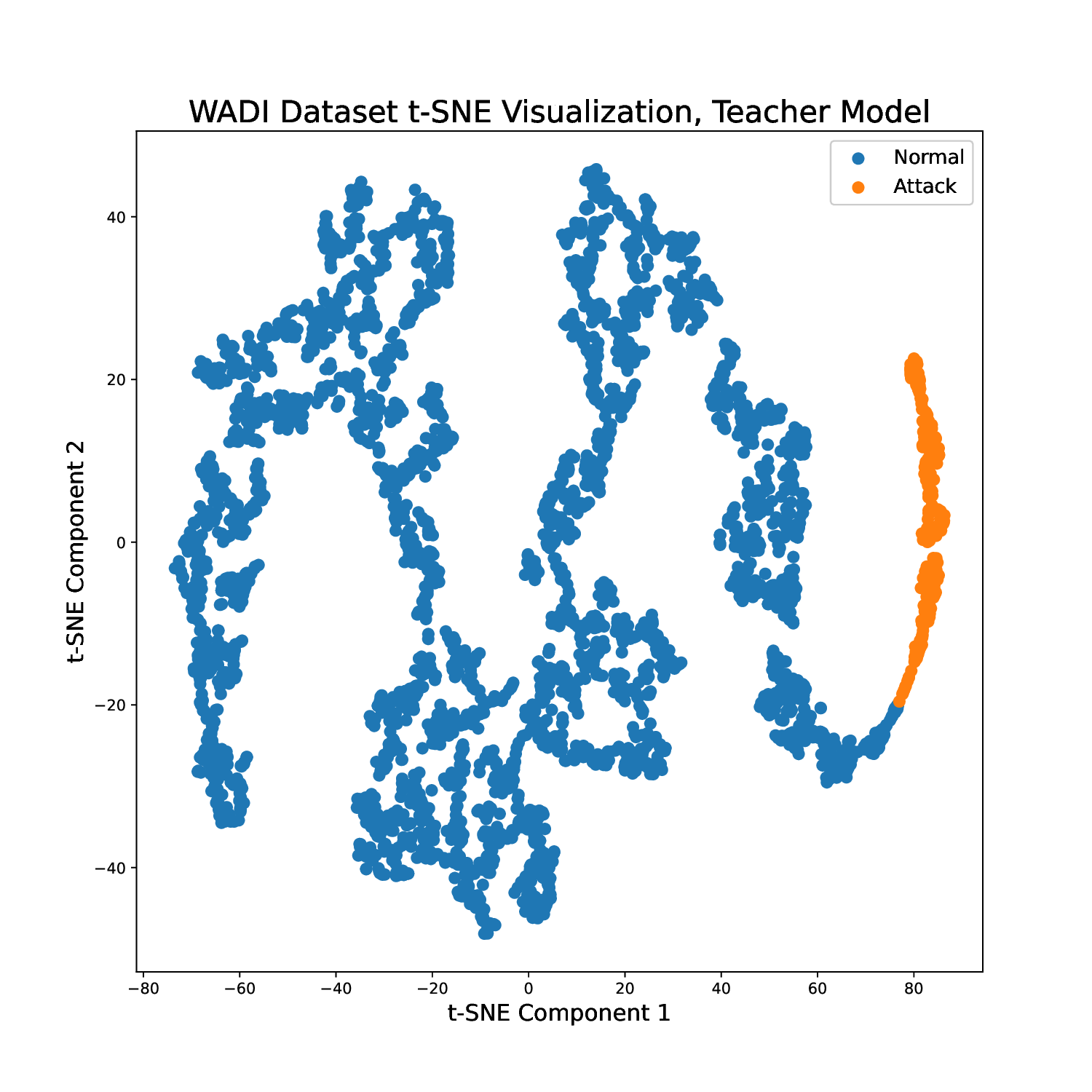} \\
        (a) SWaT KAN (Teacher)  & (b) WADI KAN (Teacher)  \\
        \includegraphics[width=0.43\textwidth]{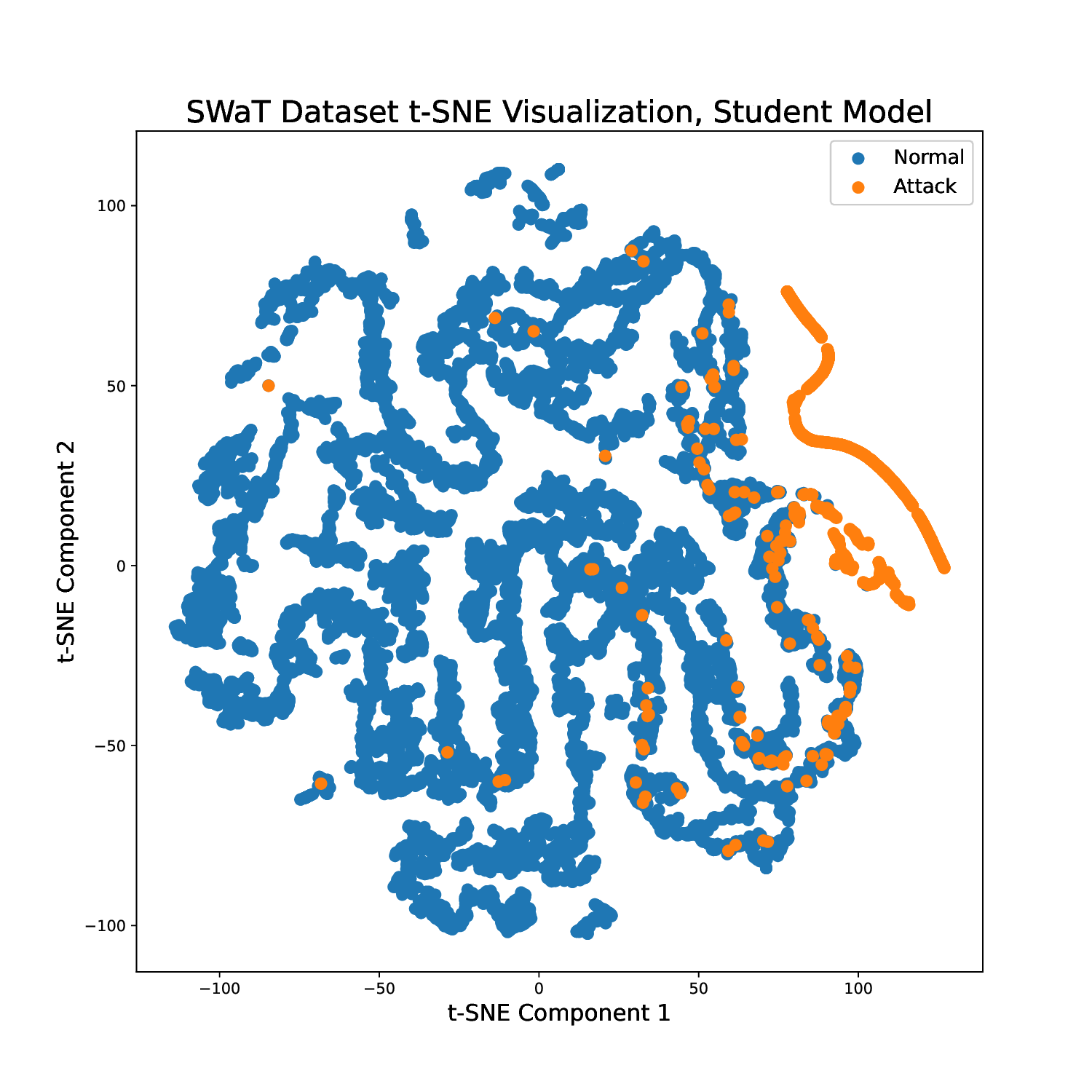} &
        \includegraphics[width=0.43\textwidth]{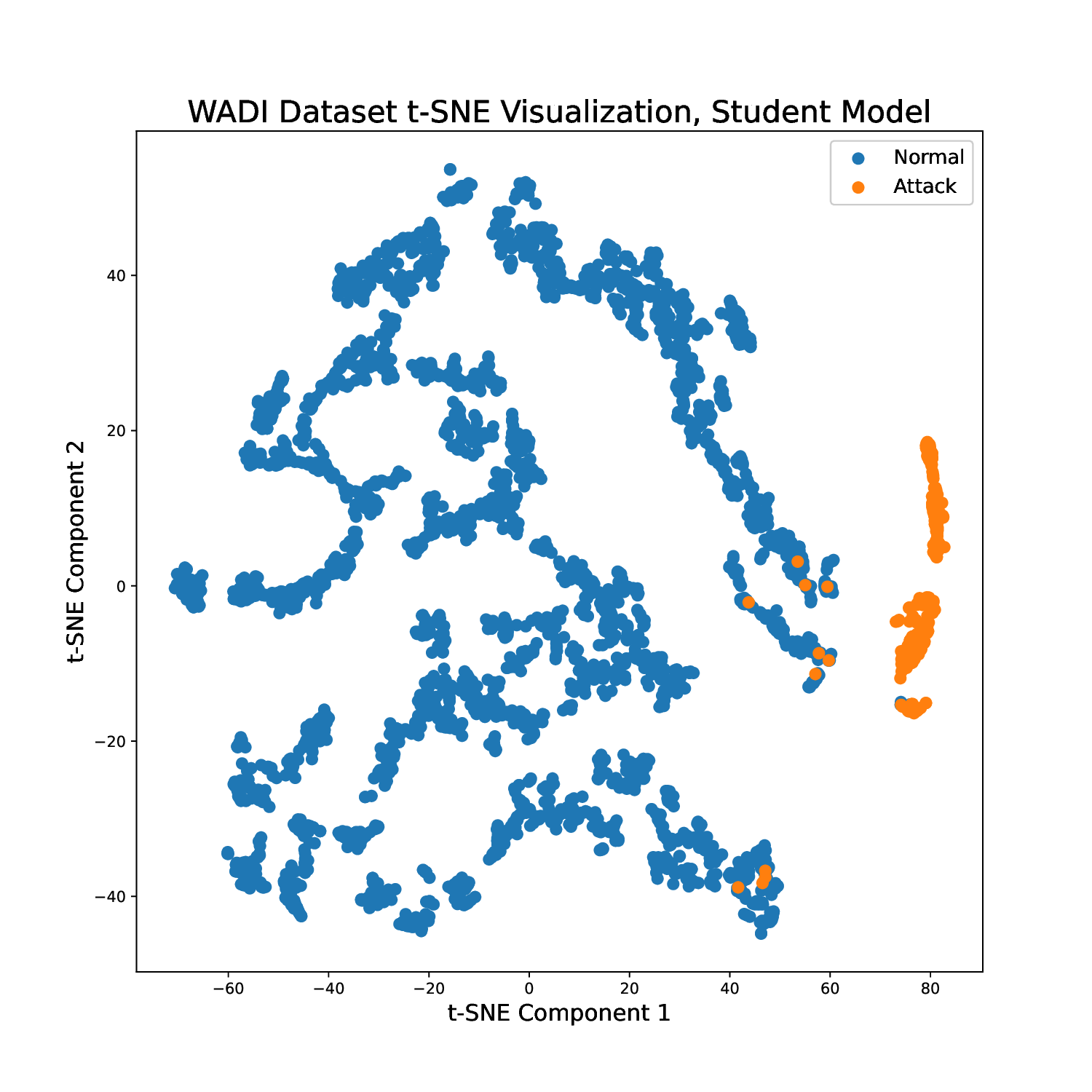} \\
        (c) SWaT MLP (Student) & (d) WADI MLP (Student) \\
        \includegraphics[width=0.43\textwidth]{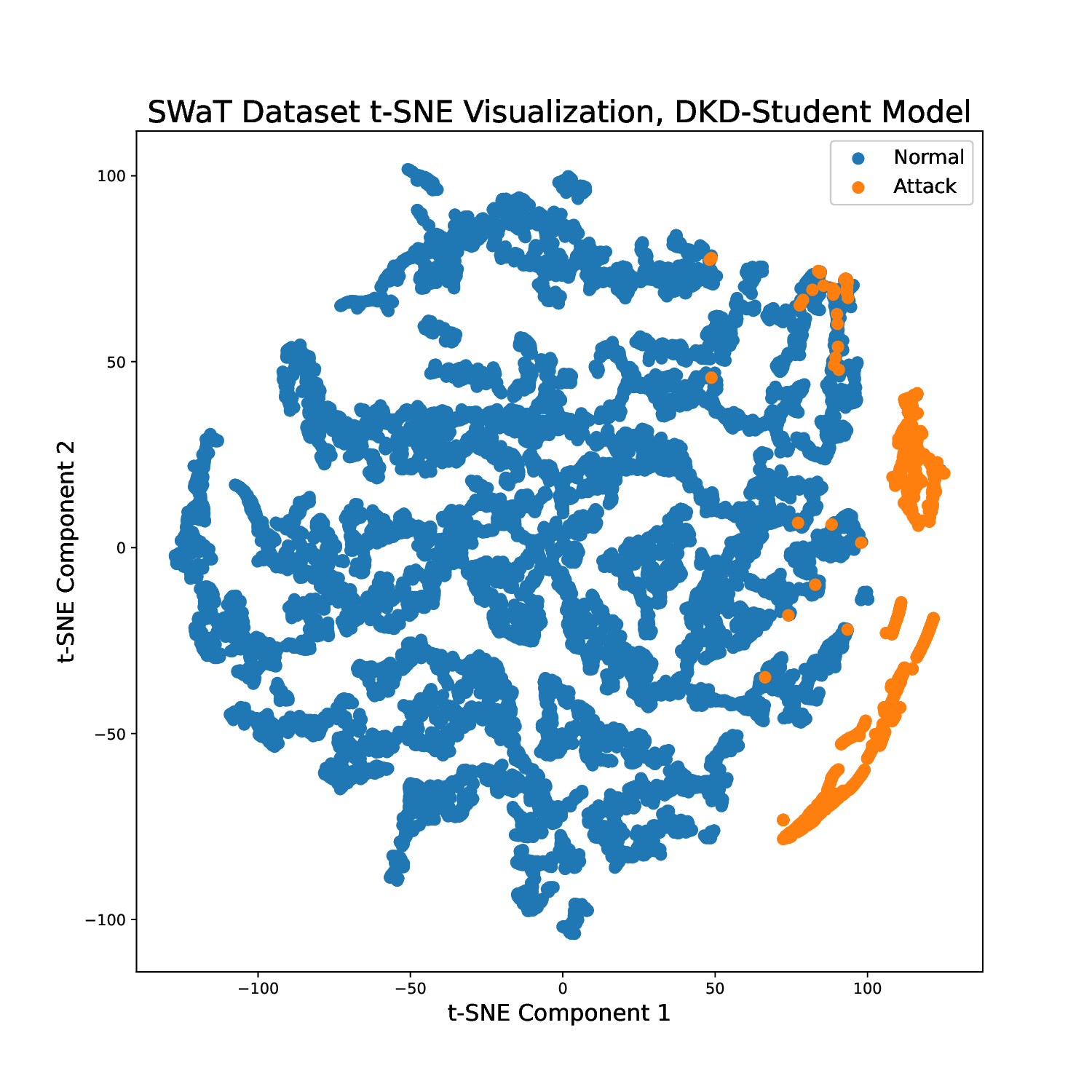} &
        \includegraphics[width=0.43\textwidth]{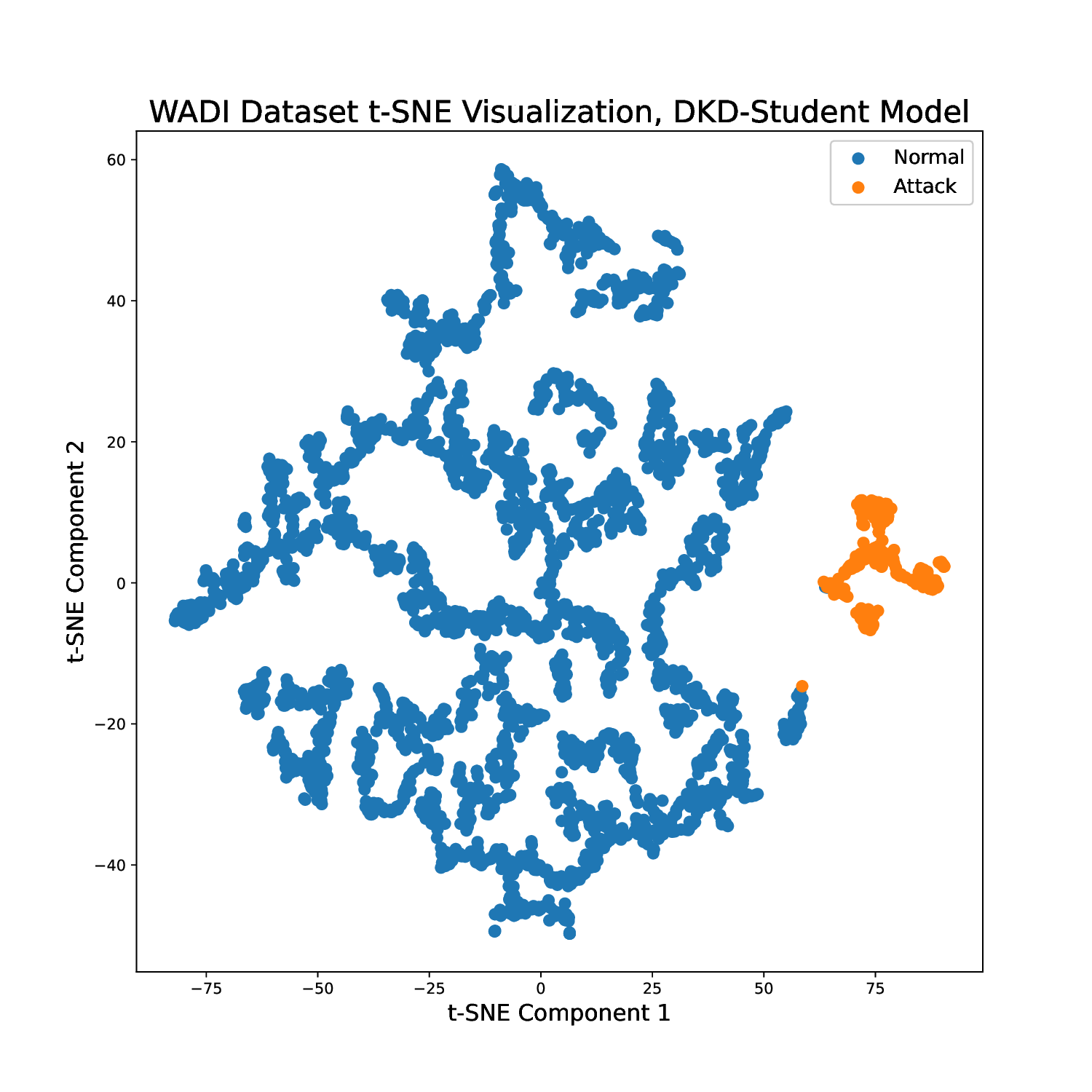} \\
        (e) SWaT DKD-MLP & (f) WADI DKD-MLP \\
    \end{tabular}
    \caption{t-SNE visualization of different models on SWaT and WADI datasets.}
    \label{fig_tsne_visualization}
\end{figure}
\begin{figure}[p]
    \centering
    \begin{tabular}{cc}
        \includegraphics[width=0.43\textwidth]{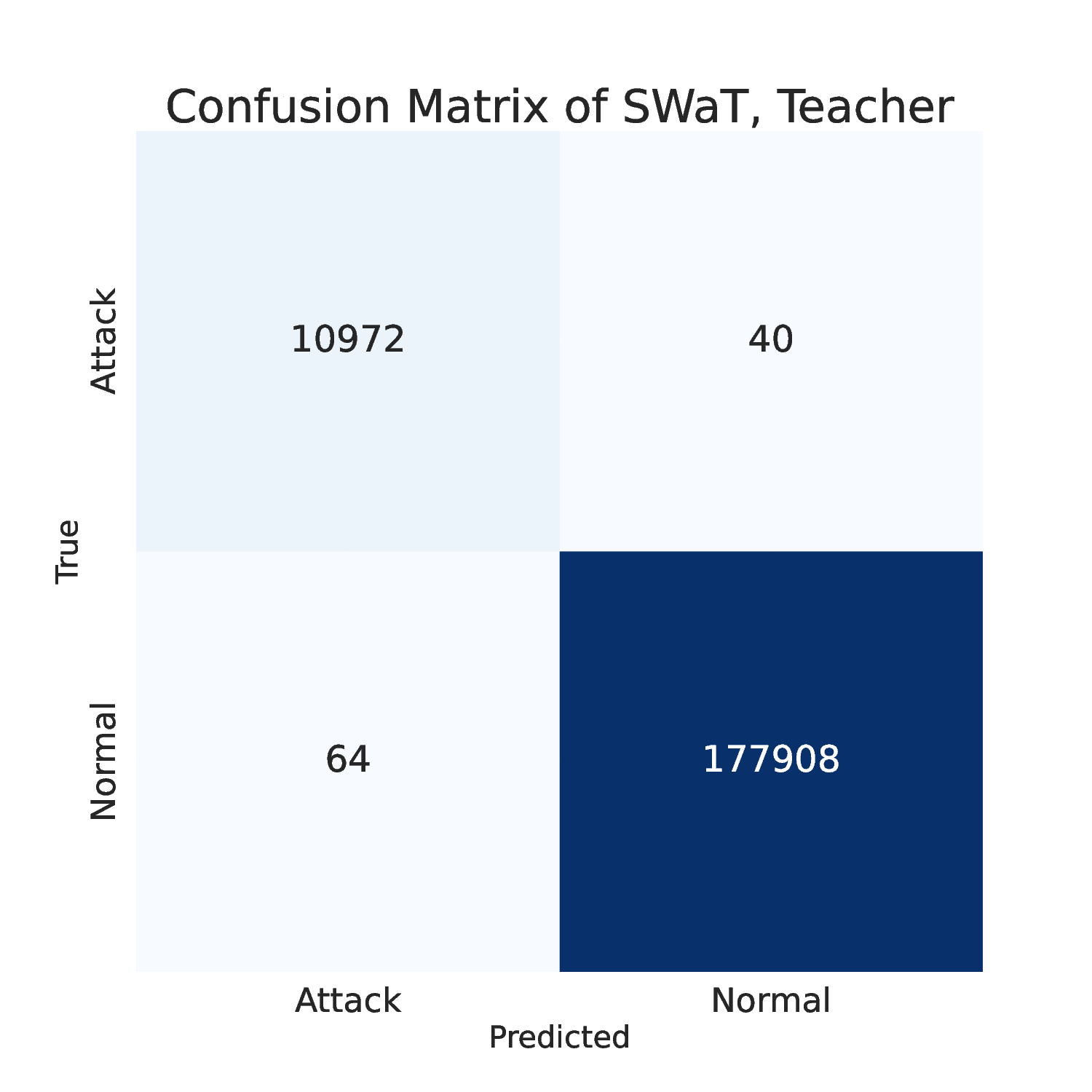} &
        \includegraphics[width=0.43\textwidth]{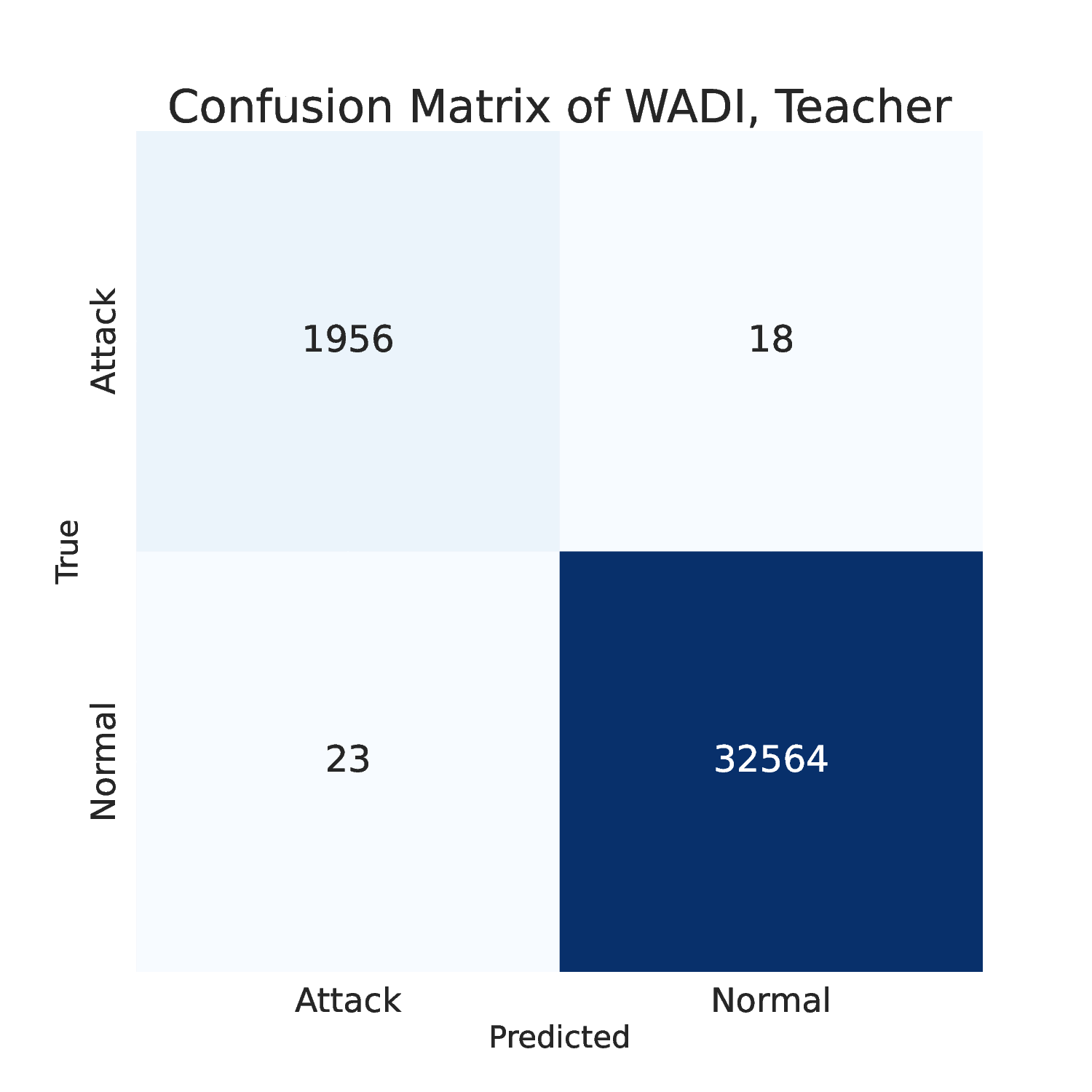} \\
        (a) SWaT KAN (Teacher) & (b) WADI KAN (Teacher) \\
        \includegraphics[width=0.43\textwidth]{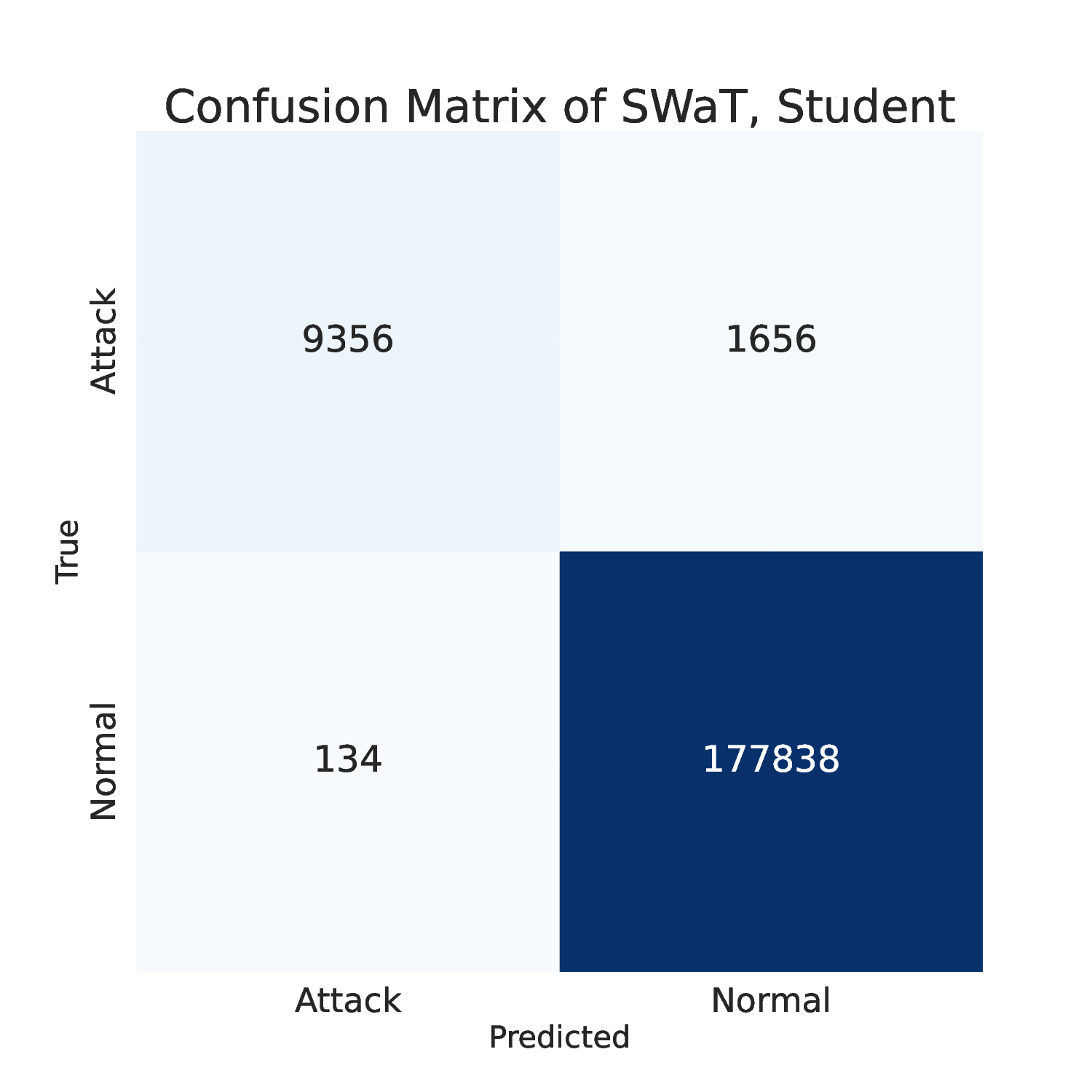} &
        \includegraphics[width=0.43\textwidth]{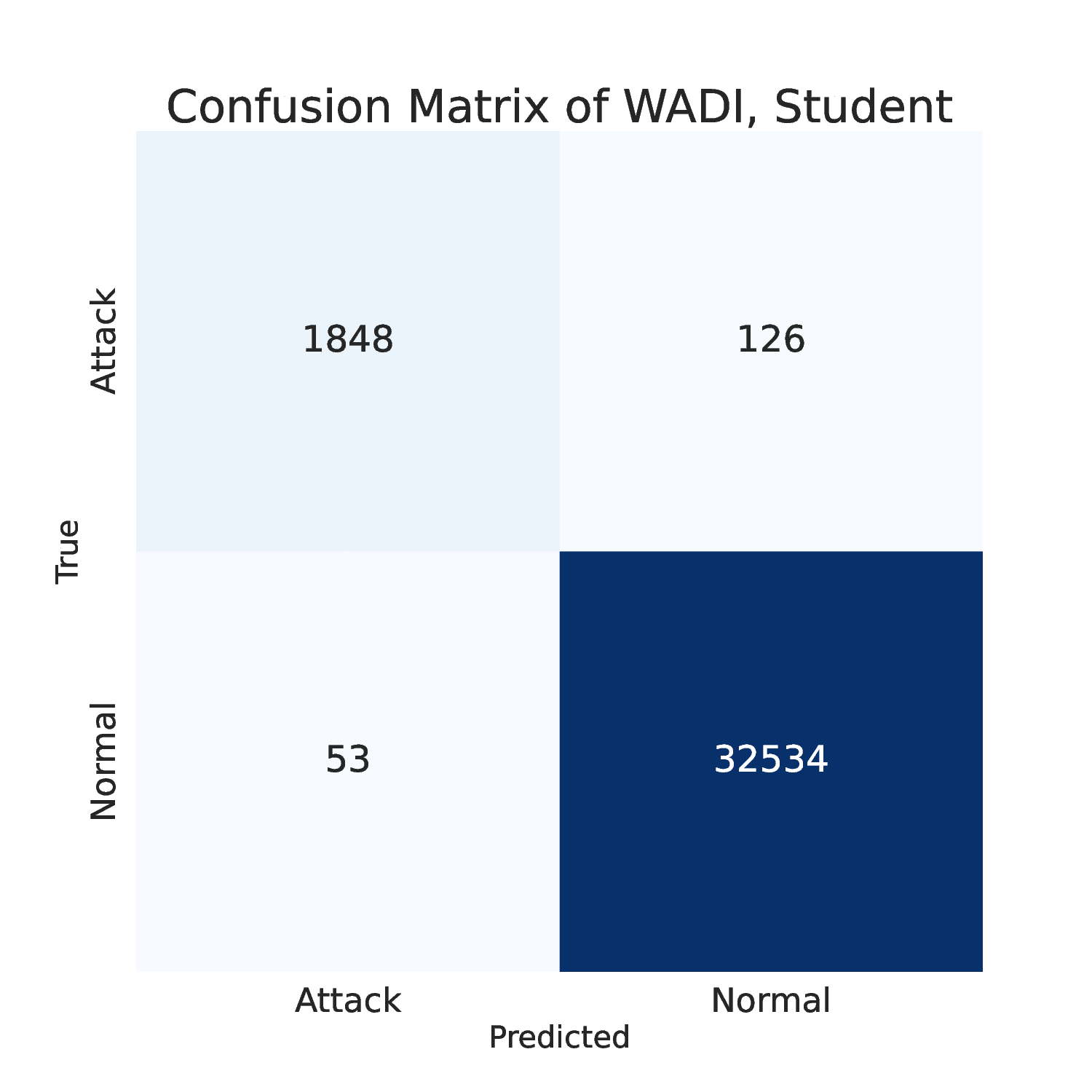} \\
        (c) SWaT MLP (Student) & (d) WADI MLP (Student) \\
        \includegraphics[width=0.43\textwidth]{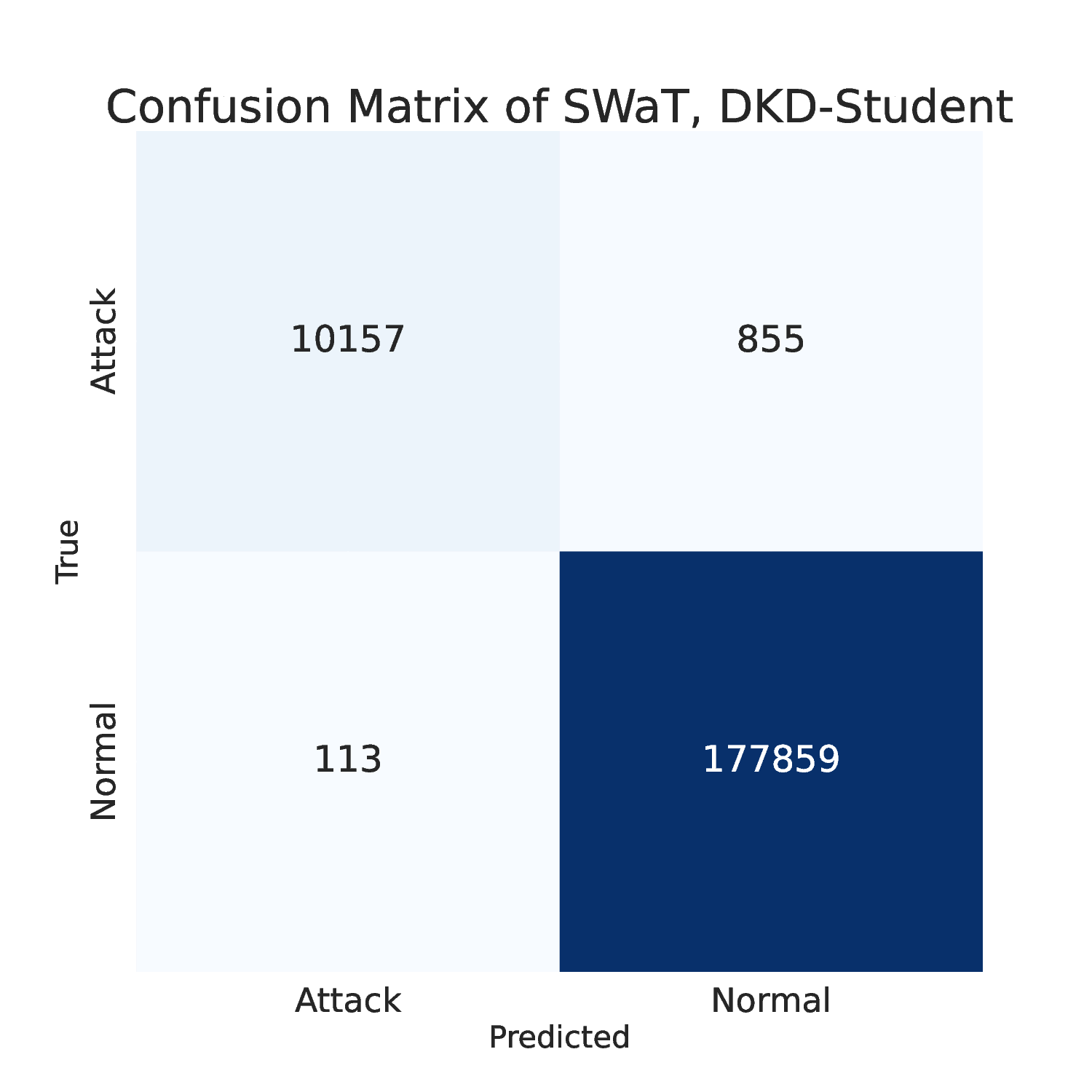} &
        \includegraphics[width=0.43\textwidth]{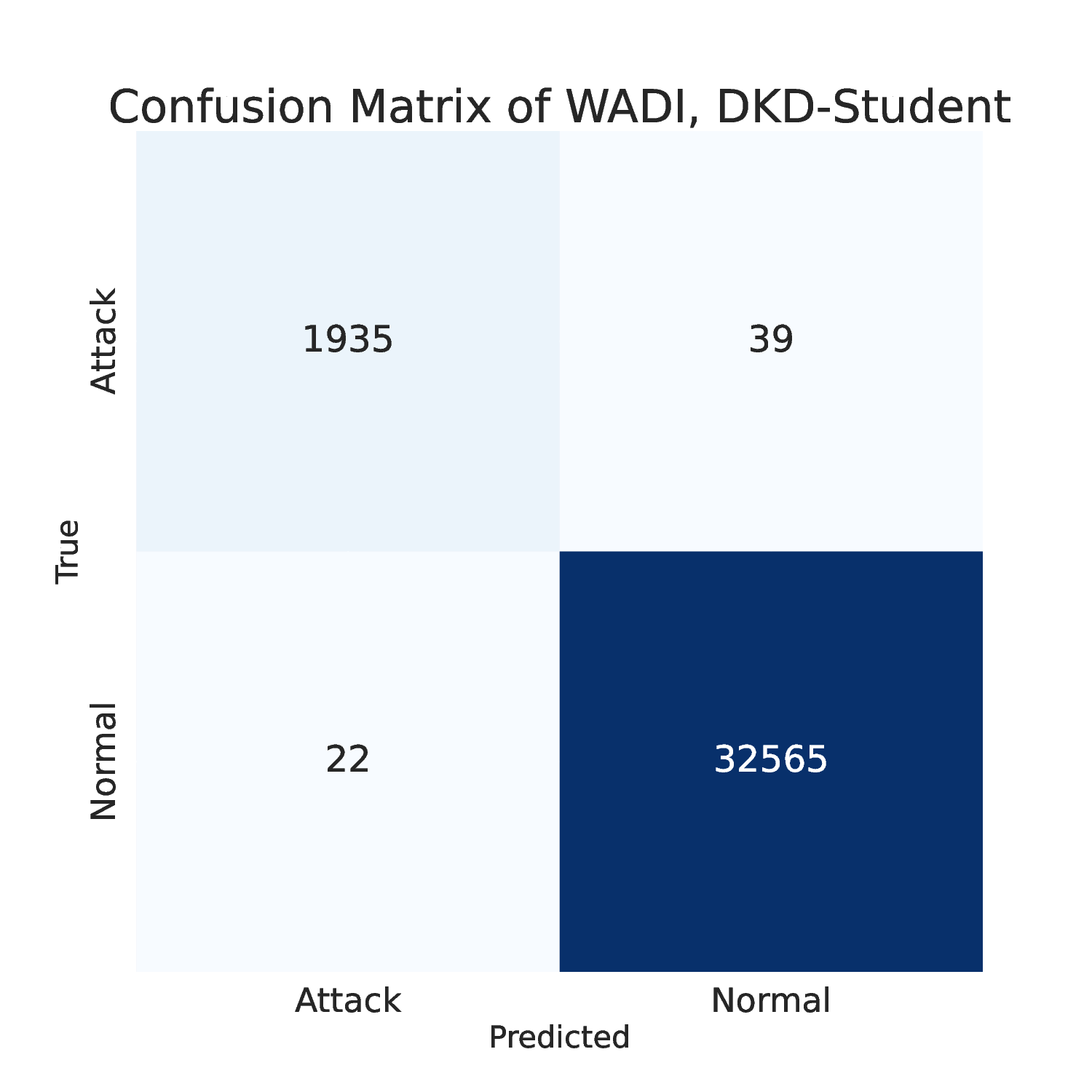} \\
        (e) SWaT DKD-MLP & (f) WADI DKD-MLP \\
    \end{tabular}
    \caption{Confusion matrices of different models on SWaT and WADI datasets.}
    \label{fig_conf_matrix}
\end{figure}

\section{Results} \label{sec_Results}
\autoref{fig_tsne_visualization} illustrates the t-SNE visualizations of the output representations produced by the MLP, KAN, and DKD-MLP models. The DKD-MLP model demonstrates superior class separability compared to the standard MLP model. Notably, for the SWaT dataset, the t-SNE patterns of the DKD-MLP model closely resemble those of the teacher model, indicating effective KD. Although the similarity is less pronounced for the WADI dataset, the clustering of attack samples remains consistent with the teacher model, highlighting the DKD-MLP model's ability to replicate the teacher's behavior in critical regions.

Confusion matrices in \autoref{fig_conf_matrix}, further evaluate the accuracy of the proposed framework. As expected, the teacher models demonstrate the highest accuracy across both datasets. However, the DKD-MLP model shows clear improvements over the baseline MLP model, reinforcing the effectiveness of the distillation-based training strategy.

\autoref{tab_swat_comparison} and \autoref{tab_wadi_comparison} present a comparative analysis between our approach and existing methods. Compared to the models proposed in~\cite{ghorbani2025using}, our method operates with substantially fewer parameters. The enhanced results reported in those studies can largely be attributed to two key factors: (i) a higher number of training epochs, and (ii) reliance on an alternative version of the WADI dataset. Despite these differences, our model achieves competitive performance in CPS intrusion detection while maintaining a lightweight architecture, which contributes to improved generalization, lower computational overhead, and faster inference.

\section{Conclusion} \label{sec_Conclusion}
This work presented a lightweight and highly efficient DKD framework for binary intrusion detection. Through extensive experiments on the SWaT and WADI datasets, we demonstrated that the proposed DKD-MLP models, trained using DKD, achieve competitive performance while drastically reducing model complexity. Specifically, our student models contain less than 2\% of the parameters of their teacher models, yet retain high F1-scores, achieving up to 95.45\% on SWaT and 98.45\% on WADI. 

The key to this success lies in the effective separation of the knowledge transfer process into target class and non-target class components, enabling fine-grained control over the distillation process via independent weighting factors $\alpha$ and $\beta$. This decoupling mitigates the limitations of classical KD where the influence of non-target knowledge diminishes for well-classified samples. In addition to performance improvements, our approach ensures scalability and deployment feasibility in resource-constrained applications, which is crucial for fast intrusion detection in industrial and critical infrastructure systems. 

While the framework performs robustly across datasets, we observed that hyperparameter tuning on the SWaT dataset is particularly challenging and warrants deeper investigation. Preliminary results suggest that adjusting parameters such as using a smaller $\lambda$ or reducing $w_{warmup}$ could potentially improve performance. However, these settings may also introduce instability during training, highlighting the need for a more systematic hyperparameter optimization approach.

\bibliographystyle{unsrt}  
\bibliography{references}  

\end{document}